\documentclass[fleqn,usenatbib]{mnras}

\usepackage{newtxtext,newtxmath}

\usepackage[T1]{fontenc}
\usepackage{caption}
\usepackage{subfigure}
\DeclareCaptionLabelFormat{continued}{#1~#2 (continued)}

\newcommand{\degree}{$^{\circ}$}

\newcommand{\logg}{$\log(g)$}

\DeclareRobustCommand{\VAN}[3]{#2}
\let\VANthebibliography\thebibliography
\def\thebibliography{\DeclareRobustCommand{\VAN}[3]{##3}\VANthebibliography}

\usepackage{graphicx}	
\usepackage{amsmath}	

\title[OGLE-BLAP-009 -- A Case Study]{OGLE-BLAP-009 -- A Case Study for the Properties and Evolution of Blue Large-Amplitude Pulsators}

\author[C. W. Bradshaw et al.]{Corey W. Bradshaw,$^{1}$\thanks{E-mail: corey.bradshaw@ttu.edu}
Matti Dorsch,$^{2,3}$
Thomas Kupfer,$^{4,1}$
Brad N. Barlow,$^{5,6}$
Uli Heber,$^{3}$
Evan B. Bauer,$^{7}$
\newauthor Lars Bildsten,$^{8}$
Jan van Roestel$^{9}$\\
$^{1}$Department of Physics and Astronomy, Texas Tech University, Lubbock, TX 79409, USA\\
$^{2}$Institut für Physik und Astronomie, Universit\"at Potsdam, 14476 Potsdam-Golm, Germany\\
$^{3}$Dr. Karl Remeis-Observatory \& ECAP, Astronomical Institute, Friedrich-Alexander University Erlangen-Nuremberg (FAU), 96049 Bamberg, Germany\\
$^{4}$Hamburger Sternwarte, University of Hamburg, Gojenbergsweg 112, 21029 Hamburg, Germany\\
$^{5}$Department of Physics and Astronomy, High Point University, High Point, NC, 27268, USA\\
$^{6}$Department of Physics and Astronomy, University of North Carolina at Chapel Hill, Chapel Hill, NC 27599, USA\\
$^{7}$Center for Astrophysics | Harvard \& Smithsonian, Cambridge, MA 02138, USA\\
$^{8}$Kavli Institute for Theoretical Physics, University of California, Santa Barbara, CA 93106, USA\\
$^{9}$Anton Pannekoek Institute for Astronomy, University of Amsterdam, 1090 GE Amsterdam, The Netherlands\\
}

\date{Accepted XXX. Received YYY; in original form ZZZ}

\pubyear{2023}

\begin{document}
\label{firstpage}
\pagerange{\pageref{firstpage}--\pageref{lastpage}}
\maketitle

\begin{abstract}
Blue large-amplitude pulsators (BLAPs) make up a rare class of hot pulsating stars with effective temperatures of $\approx$30,000~K and surface gravities of 4.0 - 5.0 dex (cgs). The evolutionary origin and current status of BLAPs is not well understood, largely based on a lack of spectroscopic observations and no available mass constraints. However, several theoretical models have been proposed that reproduce their observed properties, including studies that identify them as pulsating helium-core pre-white dwarfs (He-core pre-WDs). We present here follow-up high-speed photometry and phase-resolved spectroscopy of one of the original 14 BLAPs, OGLE-BLAP-009, discovered during the Optical Gravitational Lensing Experiment. We aim to explore its pulsation characteristics and determine stellar properties such as mass and radius in order to test the consistency of these results with He-core pre-WD models. Using the mean atmospheric parameters found using spectroscopy, we fit a spectral energy distribution to obtain a preliminary estimate of the radius, luminosity and mass by making use of the \textit{Gaia} parallax. We then compare the consistency of these results to He-core pre-WD models generated using MESA, with predicted pulsation periods implemented using GYRE. We find that our mass constraints are in agreement with a low-mass He-core pre-WD of $\approx$0.30~M$_{\odot}$. 
 
\end{abstract}



\section{Introduction}\label{sec:Introduction}
Blue large-amplitude pulsators (BLAPs) comprise a class of hot variable stars that reside in a region of the Hertzsprung Russell (HR) diagram, below the main sequence, that is typically populated with hot subdwarf stars (sdOBs). BLAPs have spectral classifications of O or B and show periodic brightness variations on the timescale of 10 to 40 minutes with amplitudes of 0.2-0.4 mag \citep{BLAPDISC2017,HD13_2022,Ramsay_2022}. Originally, 14 of these objects were discovered as a part of the Optical Gravitational Lensing Experiment (OGLE) survey of the Galactic bulge and disk \citep{OGLE4}. These stars have light curves similar to those found in fundamental mode pulsating Cepheid and RR-Lyrae stars but with higher effective temperatures ($T_{\textrm{eff}}$) of $\approx$30,000~K and surface gravities ($\log(g)$) of $\approx$4.5 \citep{BLAPDISC2017}. Since then, 4 objects were found from the Zwicky Transient Facility (ZTF) at low Galactic latitudes that exhibit large amplitude pulsations with similar $T_{\textrm{eff}}$ but higher $\log(g)$ of $\approx$5.4 and shorter pulsation periods in the range 3 to 8 minutes. These objects were then classified as high-gravity BLAPs \citep{Kupfer_2019}. \citet{Ramsay_2022} reported the discovery of 4 BLAPs as part of the OmegaWhite survey, one of which is OGLE-BLAP-009, confirming its previously observed parameters. All of their BLAPs have properties consistent with those reported by \citet{BLAPDISC2017}. Additionally, \citet{HD13_2022} and \citet{tmts_blap} each confirmed a BLAP as component of a binary pair. These discoveries constitute the only two BLAPs to be found in a binary system. More BLAP candidates have been proposed by \citet{McWhirter_2022} through cross-matching sources meeting color-corrected selection criteria in \textit{Gaia} 
DR2 \citep{gaia,Gaia_2018} to light curves from ZTF DR3. The position of BLAPs and high-gravity BLAPs on the HR diagram and corresponding effective temperatures are similar to that of B-type hot-subdwarf (sdB) stars but with lower surface gravities in the case of BLAPs (see \citealt{Heber_1986,Heber_2009,Heber_2016}).


The evolutionary history of these objects is not fully understood, and several scenarios have been proposed. Along with the initial discovery of BLAPs, \citet{BLAPDISC2017} suggested two models that may reproduce their observed properties. The first is that of a star in the helium-core (He-core) burning phase with an inflated hydrogen envelope and a mass of $\approx$$1$~M$_{\odot}$, which may occur in hot stars as a result of significant mass loss of about 75\%. The second case is that of a stripped red giant, whose energy is produced by a hydrogen-burning shell above a degenerate helium-core resulting in a pre-white dwarf (He-core pre-WD) of $\approx$$0.30$~M$_{\odot}$. Both scenarios require significant mass loss, which establishes a connection to sdB stars; however, this requirement is less extensive in the second case and was considered to be more likely \citep{BLAPDISC2017}. 

Several theoretical investigations into pulsating He-core pre-WDs have been performed. \citet{cors_2016} found that non-radial $p$, $g$ and $p-g$ mixed pulsations can occur in He-core pre-WDs due to the $\kappa - \gamma$ mechanisms. Expanding on this, \citet{Romero_2018} and \citet{cors_ev} found that evolutionary tracks of the hot counterparts to these pre-extremely-low-mass (pre-ELM) white dwarfs with masses between $0.27-0.37$~M$_{\odot}$ can reproduce the observed $T_{\textrm{eff}}$ and $\log(g)$ of BLAPs. The pulsation periods of these models were best described by low-order radial-modes or high-order $g$-mode pulsations due to the $\kappa$-mechanism. \citet{Bryne_2018, Byrne_2020} reported that the fundamental radial mode can be excited when including radiative levitation of iron-group elements in post-common envelope evolution models that result in the formation of a low-mass He-core pre-WD of $\approx$$0.31$~M$_{\odot}$ pulsating as a result of the $\kappa$-mechanism.  

\citet{Kupfer_2019}, in the initial discovery of high-gravity BLAPs, tested their observed properties with models developed using the Modules for Experiments in Stellar Astrophysics (MESA) stellar evolution code for both He-core pre-WDs as well as low-mass He-core burning stars with thin hydrogen envelopes of various masses. These tracks also included adiabatic pulsation periods calculated for the fundamental and first-overtone radial-modes. They found that the $T_{\textrm{eff}}$ and $\log(g)$ of these stars could be represented by both models, but the observed periods more closely resembled He-core pre-WDs pulsating in the fundamental radial-mode. Using the dynamical frequency for these models in order to calculate radius and mass, they found that the four high-gravity BLAPs had masses between $0.19 - 0.35$~M$_{\odot}$ with radii of $0.10 - 0.22$~R$_{\odot}$. They concluded that their objects favored the low-mass He-core pre-WD models over the more massive He-core burning models \citep{Kupfer_2019}. 

The proposed evolutionary status of a pre-WD leads to the possibility of a binary evolution channel being responsible for the production of BLAPs. However, only two BLAPs have been discovered in a binary in the case of HD 133729 and TMTS-BLAP1 \citep{HD13_2022, tmts_blap}. \citet{HD13_2022} identified a BLAP as the secondary component in the binary system HD 133729, with a late B-type main sequence primary and a 23.08433 day orbital period. \citet{tmts_blap} discovered a BLAP (TMTS-BLAP1) in an approximately 1500 day orbital period around a dwarf companion using light travel time effects. A binary population synthesis study performed by \citet{Byrne_2021} found that BLAPs can in fact be produced through mass stripping of their progenitor via common envelope evolution (CEE) or stable Roche lobe overflow (RLOF). The lack of BLAPs without observed binary companions is therefore puzzling. \citet{Meng_2020} showed that He-core burning survivors of Type Ia supernovae with masses of $\approx$0.75~M$_{\odot}$ can reproduce the observed properties of BLAPs, leading to a possible explanation for the lack of observable companions. These proposed evolutionary pathways suggest that BLAPs may serve an important role in investigating stages of binary evolution.

Hot subdwarf models for He-core and He-shell burning stars have also been studied as a potential evolutionary scenario for BLAPs. \citet{CHeB_ev} found that He-core burning subdwarfs with masses in the range of $0.70-1.10$~M$_{\odot}$ reproduced observed spectroscopic properties as well as reported rates of period change. This range of masses is consistent with findings from \citet{Meng_2020}. Investigations carried out by \citet{SHeB_ev} showed that He-shell burning stars with core masses in the range of $0.45-0.50$~M$_{\odot}$ can reproduce observed BLAP properties while these stars with higher core masses of $0.75-1.0$~M$_{\odot}$ were unable to reproduce these properties.   

It remains an open question as to which evolutionary state BLAPs currently reside in and if they all share similar origins. Currently there are no mass or radius estimates for any of the known BLAPs, excluding the high-gravity BLAPs. Additional observations that aid in constraining the physical properties of these stars will help to determine their proper status. We present a detailed follow-up analysis of one particularly interesting BLAP, OGLE-BLAP-009. This object was initially reported by \citet{Macfarlane_2017} as a "peculiar $\delta$~Scuti-type pulsating star found in the OmegaWhite survey and then independently found and formally classified as a BLAP as part of the OGLE survey of the Galactic bulge and disk. This star has a reported period of 31.94 minutes from OGLE photometry with a $T_{\textrm{eff}}$ of $31,800\pm1400$~K and a $\log(g)$ of $4.40 \pm 0.18$ from long-exposure (300~$s$) low-resolution ($R\approx800$) spectroscopy obtained using Gemini \citep{BLAPDISC2017}. In this study, we aim to estimate the fundamental properties of this star and constrain its evolutionary history. Data from time-series photometry and spectroscopy were used to derive a precise period, $T_{\textrm{eff}}$ and $\log(g)$ as well as investigate the stellar pulsations and search for radial velocity signatures of a binary companion. To obtain an initial estimate of the mass and radius we fit a spectral energy distribution and combine the results with the \textit{Gaia} EDR3 parallax measurement. We then compare these results to He-core pre-WD models developed using MESA and derive a mass and radius using a dynamical frequency calculation from GYRE predicted pulsation periods, to test this evolutionary channel. 

\section{Observations}\label{sec:Observations}
Follow up high-speed photometry was performed using the 2.1 meter (82 inch) Otto Struve Telescope with the ProEM frame-transfer CCD detector located at McDonald Observatory in Fort Davis, Texas. Data was taken over three nights in July 2021 in both $g$ and $r$-bands using 8-second exposure times, resulting in 1100 $g$-band exposures and 891 $r$-band exposures. Data reduction was performed using standard \texttt{IRAF} pipelines. All frames were bias-subtracted and flat-fielded.

Phase-resolved spectroscopy was obtained using the Keck telescopes in Hawaii. The Echellette Spectrograph and Imager (ESI; \citealt{ESI}) was used to obtain fifteen 2-minute exposures at a medium resolution ($R\approx6000$) in July 2020 covering 1.18 pulsation cycles. An additional eight 2-minute exposures were taken in June 2021 which covered 0.64 pulsation cycles. The spectra were reduced using the \texttt{MAKEE}\footnote{\url{https://sites.astro.caltech.edu/~tb/makee/}} pipeline following the standard procedure: bias subtraction, flat fielding, sky subtraction, order extraction, and wavelength calibration. 

The Low Resolution Imaging Spectrometer (LRIS; \citealt{LRIS}) was also used to collect sixteen 2-minute exposures at a low resolution ($R\approx1000$) in September 2021 covering 1.23 pulsation cycles. Table~\ref{tab:tab1} lists all the data used in this analysis. Data reduction was performed with the Lpipe pipeline\footnote{\url{https://sites.astro.caltech.edu/\~dperley/programs/lpipe.html}}\citep{per19}. 

\begin{table}
	\centering
	\caption{Complete list of data used in the analysis, including both time-series photometry and spectroscopy.}
	\label{tab:tab1}
	\begin{tabular}{lcccr} 
		\hline
		\hline
		Photometry & Date & Filter & Exp. time [s] & Exps.\\
		\hline
		OGLE 3 \& 4 & 2001-2016 & $V$, $I$ & - & 2448\\
		2.1m/Pro-EM & July 9, 2021 & $g$' & 8 & 1100\\
		2.1m/Pro-EM & July 10, 2021 & $r$' & 8 & 644\\
		2.1m/Pro-EM & July 11, 2021 & $r$' & 8 & 247\\
		\hline
		\hline
		Spectroscopy & Date & Camera & Exp. time [s] & Exps.\\
		\hline
		Keck/ESI & July 22, 2020 & - & 120 & 14\\
		Keck/ESI & June 7, 2021 & - & 120 & 8\\
		Keck/LRIS & Sept. 11, 2021 & Blue & 120 & 16\\
		\hline
	\end{tabular}
\end{table}

\section{Results}\label{sec:Results}
\subsection{Photometry} \label{sec:Photometry}
Data released from the OGLE 3 \& 4 surveys in both $V$ and $I$ bands, along with high-speed follow-up data obtained from McDonald observatory in the $g$ and $r$-bands was used in our photometric study of OGLE-BLAP-009. Combining these data together resulted in photometry covering a 20 year time-span. Using the Period04 module \citep{P04} a Fourier analysis revealed a precise period of 31.93526166(4) minutes. The Fourier spectrum showed a single dominant frequency with several harmonics at integer multiples of this value as a result of the non-sinusoidal waveform. No additional signals were found after subtracting this fundamental frequency, confirming the mono-periodicity found by both \citet{BLAPDISC2017} and additional follow-up photometry from \citet{monoperiod}. High-cadence data collected in both the $g$ and $r$-bands from McDonald Observatory were combined and their Fourier spectrum revealed a fundamental period of $31.935\pm0.002$ minutes. Calculating the individual relative flux amplitudes of each light curve resulted in $0.248\pm0.003$\,mag for the $g$-band and $0.224\pm0.003$\,mag for the $r$-band, with higher amplitudes occurring in the $g$-band, as expected for radial mode pulsators. After fitting and subtracting this dominant frequency, no additional signals were found. We also see no photometric indications of a binary companion in the form of eclipses or amplitude modulation from a reflection effect or ellipsoidal modulations. Due to the short baseline of our follow-up photometry, we do not achieve a precision capable of measuring a rate of period change in relation to previous OGLE data that is on the order of those found by \citet{BLAPDISC2017} ($10^{-7}$~yr$^{-1}$).

\begin{figure}
	\includegraphics[width=\columnwidth]{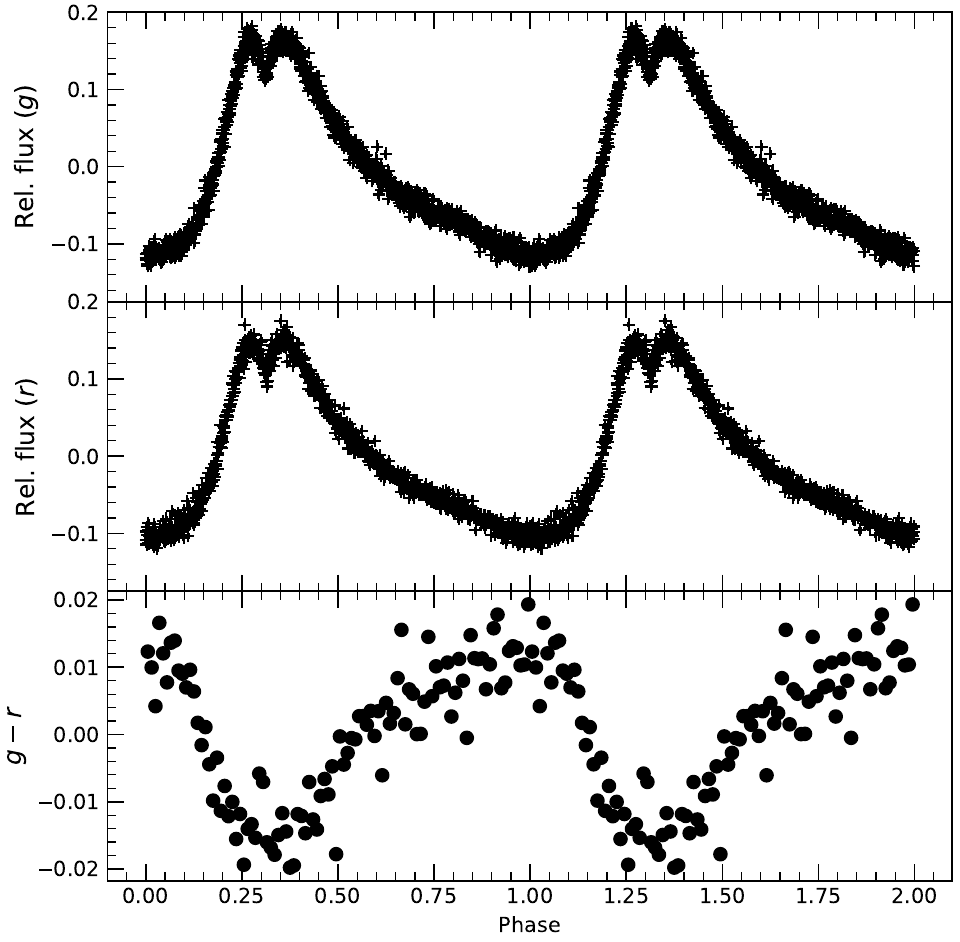} 
    \caption{Follow up photometry collected using McDonald Observatory. Top panel: $g$-band light curve, Middle panel: $r$-band light curve, Bottom panel: $g-r$ relative color variations. All are phase-folded and repeated over two pulsation cycles.}
    \label{fig:photometry}
\end{figure}
Fig.~\ref{fig:photometry} shows the light curves in the $g$ and $r$-bands in the top and middle panels, phase-folded on the calculated period of 31.93526166(4) minutes and plotted over two pulsation cycles for visualization, where phase zero is defined from the ephemeris with $t_{0}=0$~days, where we use MBJD for timing measurements. The light curves have a typical sawtooth shape with a fast rise to maximum flux covering $\approx$27\% of the pulsation cycle followed by a slow decay lasting for $\approx$63\% of the cycle. The drop to flux minimum is stalled for $\approx$10\% of the pulsation cycle, between rise and decay, where the flux makes a small "dip" about its maximum. The light curves from both \citet{BLAPDISC2017} and \citet{monoperiod} also show this double-peak feature that is clearly illustrated here in our high-cadence data. There are other BLAPs that show this feature at or near maximum flux as well (see \citealt{HD13_2022, Ramsay_2022}) and its potential origins are briefly discussed in Sec.~\ref{sec:Stellar_Pulsations}. Additionally, $g-r$ color variability is plotted in the bottom panel, showing that the flux variations are caused by stellar pulsations with bluer color (i.e., higher temperature) corresponding to the phase of maximum light. 

\subsection{Spectroscopy} \label{sec:Spectroscopy}
\begin{figure}
	\includegraphics[width=\columnwidth]{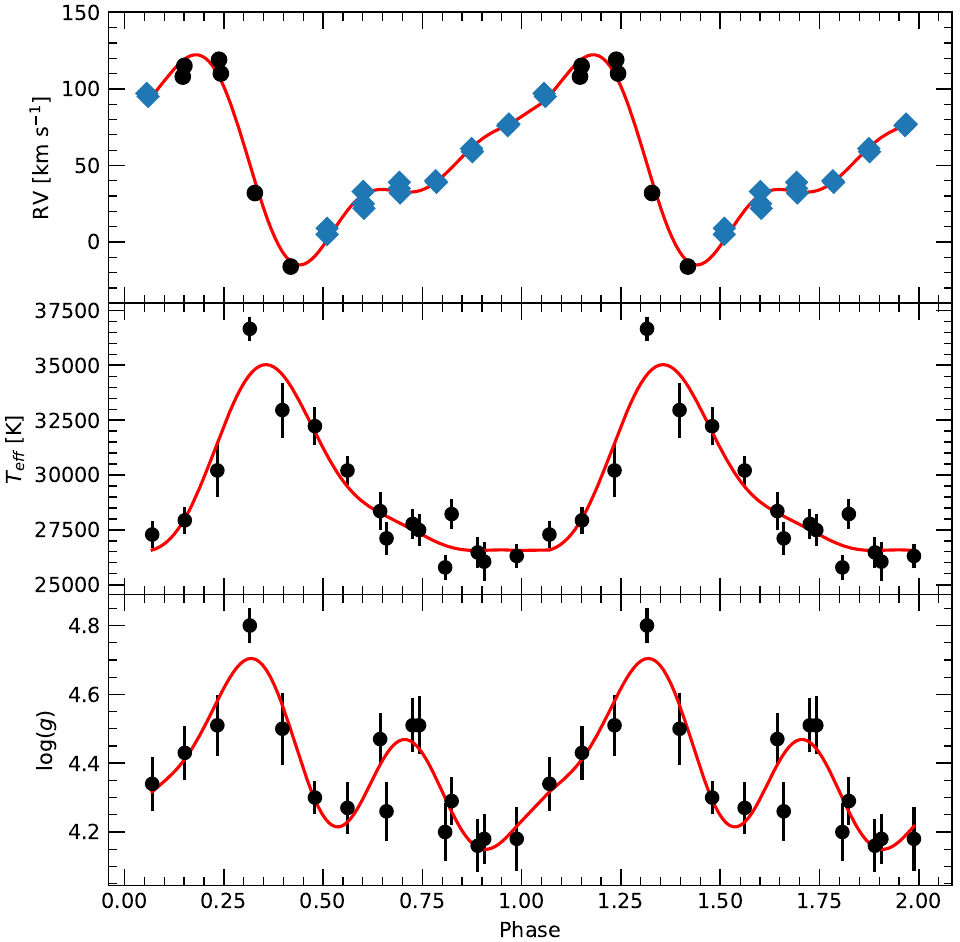}
    \caption{Results from time-series spectroscopy. Top panel: $RV$ measurements from Keck/ESI data sets. Points denoted by blue diamonds represent spectra used in metal abundance calculations. Middle panel: $T_{\textrm{eff}}$ measurements from Keck/LRIS data. Bottom panel: $\log(g)$ measurements from Keck/LRIS data. All curves are phase-folded and repeated over two pulsation cycles, where the red curves are the third order Fourier fitting (Eq.~\ref{eq1}) for each.} 
    \label{fig:spectroscopy}
\end{figure}

The large photometric variability in the photosphere of this star emphasizes the need for time-series spectroscopy to better understand the atmospheric properties of BLAPs, as they are expected to be highly phase-dependent. Time-series spectroscopy collected from Keck/ESI and Keck/LRIS show large $RV$, $T_{\textrm{eff}}$ and $\log(g)$ variability over the pulsation cycle as typically seen in radial-mode pulsators. The large variability in these parameters is caused by physical changes in the star's photosphere such as acceleration, radius expansion and contraction, and temperature changes. 

For radial velocity ($RV$) measurements, data collected from Keck/ESI was used, which provides high stability throughout observations, making it the ideal instrument for accurate phase-resolved $RV$ measurements. These were performed by fitting Gaussians, Lorentzians, and polynomials to the hydrogen and helium lines to cover continuum, line, and line core of the individual absorption features using the \texttt{FITSB2} routine \citep{nap04a}. The procedure is described in full detail in \citet{gei11a}. We fitted the wavelength shifts and compared their measurement to the rest wavelengths using a $\chi^2$-minimization. High-resolution echelle spectra are not well suited to measure $T_{\textrm{eff}}$ and $\log(g)$ because the broad hydrogen absorption lines span several individual echelle orders and merging of the echelle spectra could introduce systematic errors. Therefore, we used the Keck/LRIS spectra for $T_{\textrm{eff}}$ and $\log(g)$ measurements. We use spectral models computed with \textsc{Tlusty}/\textsc{Synspec} \citep{hub17} covering helium dominated atmospheres. The approach and the models are described in detail in \citet{Dorsch2021}. This model grid uses Fe and Ni abundances of 1.5 times solar and 10 times solar, respectively. Using the spectral modelling tool SPAS \citep{spas}, the individual Keck/LRIS spectra were fitted for $T_{\textrm{eff}}$, $\log(g)$, and helium abundance ($\log(y)=\log(n(\mathrm{He})/n(\mathrm{H})$) using atmospheric models. The phase resolved results for $\log(y)$ show no clear variability due to the large uncertainty from fitting individual spectra (see Table~\ref{tab:tabB1}). We instead perform a full metal abundance analysis using the co-added ESI spectra in Sec.~\ref{sec:Metal Abundance}. Fig.~\ref{fig:spectroscopy} shows the resulting phase-folded spectroscopic measurements. 
 
In order to uncover the mean values from the resulting $RV$,  $T_{\textrm{eff}}$ and $\log(g)$ curves, it was necessary to fit a model that best represented their non-sinusoidal waveforms. A third order ($n=3$) Fourier expansion (Eq.~\ref{eq1}) was found to be the best fit to each of the phase-folded curves. 
\begin{equation} \label{eq1}
    RV = a_{0}+\sum_{n=1}^3 a_{n}\sin(2 \pi n x) + b_{n}\cos(2 \pi n x)
\end{equation}
Where $x$ represents the phase at each measurement. We take $a_{0}$ from this model as the mean about which the variability occurs and report these as the average $\langle RV \rangle$, $\langle T_{\textrm{eff}}\rangle$ and $\langle\log(g)\rangle$ of the star. Table~\ref{tab:coeff} contains a list of the best fitting coefficients found for each data set. 
\begin{table}
\centering
\caption{List of best fitting Fourier coefficients derived for $RV$, $T_{\textrm{eff}}$, and {\logg}}
\label{tab:coeff}
\begin{tabular}{lrcr}
		\hline
		\hline
Coefficient & $RV$~[km~s$^{-1}$] & $T_{\textrm{eff}}$~[K] & $\log(g[\textrm{cm~s}^{-2}])$ \\
		\hline
$a_{0}$ & $53.36 \pm 1.44$ & $29,315.33 \pm 331.36$ & $4.38 \pm 0.02$\\
$a_{1}$ & $20.83 \pm 2.01$ & $2508.77 \pm 448.43$ & $0.13 \pm 0.03$\\
$a_{2}$ & $19.87 \pm 2.07$ & $-1433.19 \pm 459.45$ & $-0.02 \pm 0.03$\\
$a_{3}$ & $-12.88 \pm 1.97$ & $36.50 \pm 459.22$ & $0.04 \pm 0.03$\\
$b_{1}$ & $44.78 \pm 2.05$ & $-2856.67 \pm 489.29$ & $-0.08 \pm 0.04$\\
$b_{2}$ & $-14.47 \pm 1.95$ & $-427.03 \pm 476.38$ & $-0.14 \pm 0.03$\\
$b_{3}$ & $-3.37 \pm 2.00$ & $513.13 \pm 461.15$ & $0.08 \pm 0.03$\\
		\hline
\end{tabular}
\end{table}

$RV$ calculations resulted in a peak-to-peak amplitude of $135\pm5$~km~s$^{-1}$ about a mean of $53.36\pm1.44$~km~s$^{-1}$ due to a large radius change throughout the pulsation cycle (see Sec.~\ref{sec:Stellar_Pulsations}). The top panel of Fig.~\ref{fig:spectroscopy} shows the result of $RV$ measurements made using both sets of Keck/ESI data taken approximately one year apart in July 2020 and June 2021. The data points represent the average $\langle RV \rangle$ of lines: H$\alpha$, H$\beta$, H$\gamma$, H$\delta$ and He{\sc I} 4026, 4388, 4472, 4922, 5876, 6678 \AA. Error bars for each point are included but small compared to the amplitude, only on the order of a few km~s$^{-1}$. Both data sets were then phase folded on the period of 31.93526166(4) minutes. Note that there is no average radial velocity variation between the two epochs as would be likely if OGLE-BLAP-009 were in a binary orbit (see Sec.~\ref{sec:Radial_Velocity_Monitoring}). 

Atmospheric fittings using data collected from Keck/LRIS show a peak-to-peak $T_{\textrm{eff}}$ variation of $10,868\pm784$~K about a mean of $29,315\pm331$~K and a peak-to-peak $\log(g)$ amplitude of $0.64\pm0.01$ about a mean of $4.38\pm0.03$ which can be seen in the middle and bottom panel of Fig.~\ref{fig:spectroscopy}, respectively. These average measurements are consistent with results reported by \citet{BLAPDISC2017}. The phasing of these atmospheric fittings is such that at peak temperature the star is at maximum luminosity and $\log(g)$. This is an indication of the $T_{\textrm{eff}}$ change as the predominant factor in flux variations seen in the light curve, which is also consistent with the phasing of the $g-r$ color variations. Note that both $T_{\textrm{eff}}$ and $\log(g)$ have a significantly higher measurement at their peak. We suspect that a shock is likely to trigger at minimum radius due to compression of the photosphere and travel outward through the outer layers of the stellar envelope, resulting in a spike of $T_{\textrm{eff}}$ and $\log(g)$ that our spectral models may fail to adequately measure. \cite{Jeffery2022} has recently shown for the example of the extreme helium star V652\,Her that hydrostatic model spectra as used here can significantly deviate from a proper dynamical treatment at these phases. 

\subsection{Metal abundance analysis} \label{sec:Metal Abundance}

\begin{figure*}
	\includegraphics[width=\textwidth]{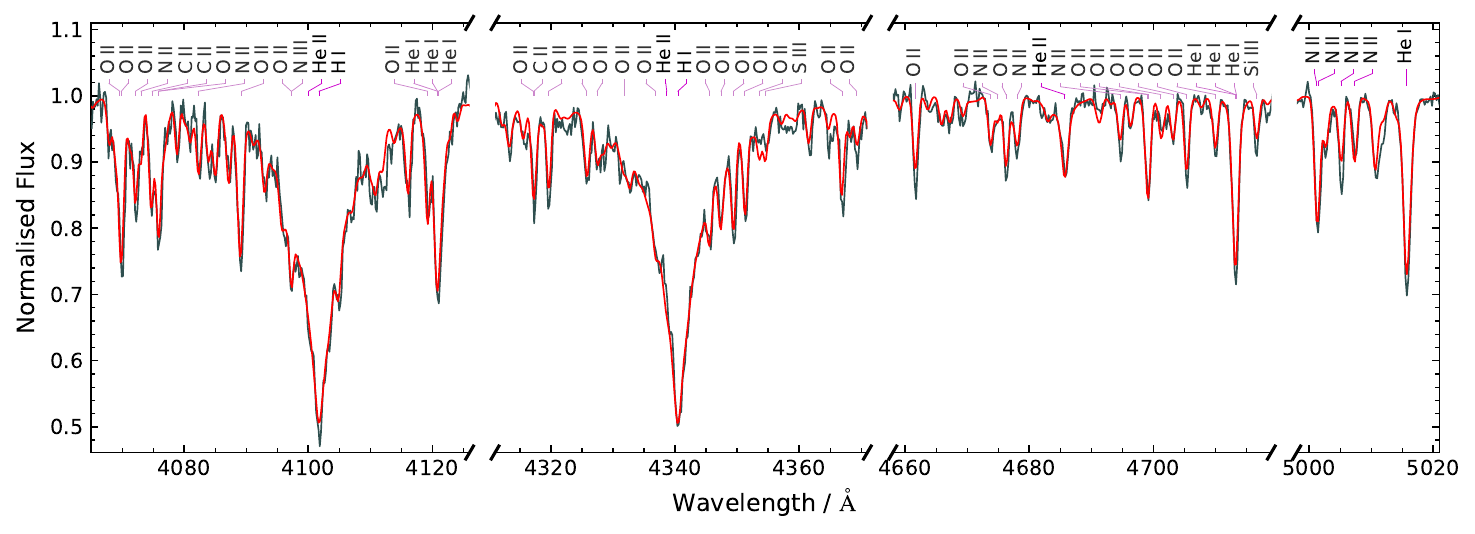}\vspace{-20pt}
	\includegraphics[width=\textwidth]{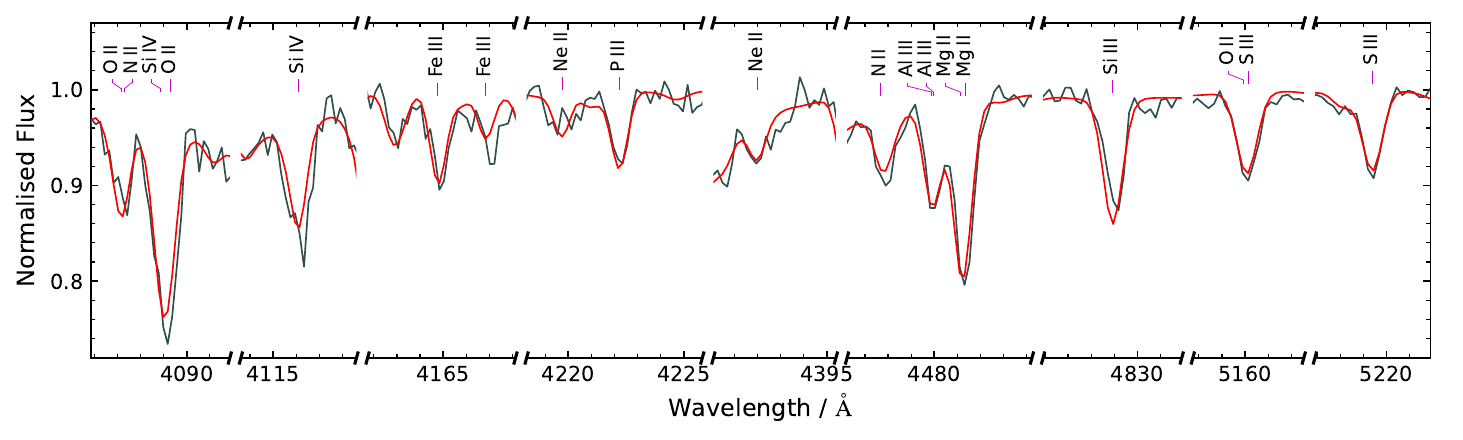}
    \caption{Example fittings of the synthetic \textsc{Tlusty}/\textsc{Synspec} model spectrum (red line) to the co-added radial-velocity corrected ESI spectra (black line) for notable metal lines. The model spectrum parameters are: $T_\textrm{eff}$ = 27,250~K, $\log(g)$ = 4.35, $\log(y)$ = -0.6, and abundances as stated in Table~\ref{tab:abundances}.}
    \label{fig:ESI_HHe}
\end{figure*}

The optical spectrum of OGLE-BLAP-009 shows unusually strong metal lines. 
In order to estimate the surface metal composition of OGLE-BLAP-009, we co-added the radial velocity-corrected ESI exposures least affected by the compression phase. The spectra used for this are marked blue in Fig.\ \ref{fig:spectroscopy}. 
We then constructed a \textsc{Tlusty}/\textsc{Synspec} model spectrum based on the previously derived low-$T_\mathrm{eff}$ phase atmospheric parameters, $\log g = 4.35$ and $\log y = -0.6$. 
Because \ion{He}{ii} 4686\,\AA\ is very sensitive to $T_\mathrm{eff}$, we used this line to estimate a best-fit value of $T_\mathrm{eff}=27,250$\,K for our metal line-blanketed model. 
The metal abundances were then iteratively adjusted to match the strong lines in the co-added ESI spectrum. 
In particular the many strong \ion{C}{ii-iii}, \ion{N}{ii-iii}, and \ion{O}{ii-iii} lines are well matched by this model, as shown in the top panel of Fig.\ \ref{fig:ESI_HHe}. 
Notable lines from heavier metals include \ion{Ne}{i} 6143, 6402\,\AA, \ion{Ne}{ii} 4392, 4409\,\AA, the \ion{Mg}{ii} 4481\,\AA\ doublet, \ion{Al}{iii} 4150, 4480, 4529\,\AA, \ion{P}{iii} 4222, 4247\,\AA, as well as many \ion{Si}{iii-iv}, \ion{S}{iii}, and \ion{Fe}{iii} lines. Examples are shown in the bottom panel of Fig.\ \ref{fig:ESI_HHe}. 
All of these metals were included in non-LTE using the largest model atoms distributed with \textsc{Tlusty} 205 \citep{hub17} with the exception of phosphorous, for which no good model atom is available. 
A full comparison between the co-added ESI spectrum and our best-fit model is shown in Fig.\ \ref{fig:full_spectrum}. 

All metal abundances turned out to be strongly enhanced when compared to solar values at factors between two and five; they are listed in Table \ref{tab:abundances}. 
The especially strong enhancement in nitrogen, about 25 times solar by number fraction, is evidence for material processed by hydrogen fusion in the CN-cycle.

\begin{table}
\centering
\caption{Surface abundances from the coadded ESI spectrum by number fraction $\varepsilon = n(\mathrm{X}) /n(\mathrm{all})$ and compared to the solar values $\varepsilon _\odot$. Phases with the highest  $T_\mathrm{eff}$ were not considered for the co-added spectrum. Both the abundances and uncertainties (68\,\%) were estimated by comparison with model spectra. 
}
\label{tab:abundances}
\begin{tabular}{lcc}
		\hline
		\hline
Element & $\log \varepsilon$ &  $\log \varepsilon /\varepsilon _\odot$ \\
		\hline
H & $-0.10\pm0.02$ & $-0.06\pm0.02$ \\
He & $-0.70\pm0.08$ & $0.41\pm0.08$ \\
C & $-2.82\pm0.10$ & $0.79\pm0.11$ \\
N & $-2.80\pm0.10$ & $1.41\pm0.11$ \\
O & $-3.03\pm0.10$ & $0.32\pm0.11$ \\
Ne & $-3.41\pm0.20$ & $0.69\pm0.23$ \\
Mg & $-3.88\pm0.10$ & $0.56\pm0.11$ \\
Al & $-5.02\pm0.10$ & $0.56\pm0.10$ \\
Si & $-3.97\pm0.15$ & $0.56\pm0.15$ \\
P & $-5.96\pm0.15$ & $0.66\pm0.15$ \\
S & $-4.65\pm0.10$ & $0.27\pm0.10$ \\
Fe & $-3.89\pm0.20$ & $0.65\pm0.20$ \\
		\hline
\end{tabular}
\end{table}

\subsection{Spectral energy distribution} \label{sec:SED}
\begin{figure}
	\includegraphics[width=\columnwidth]{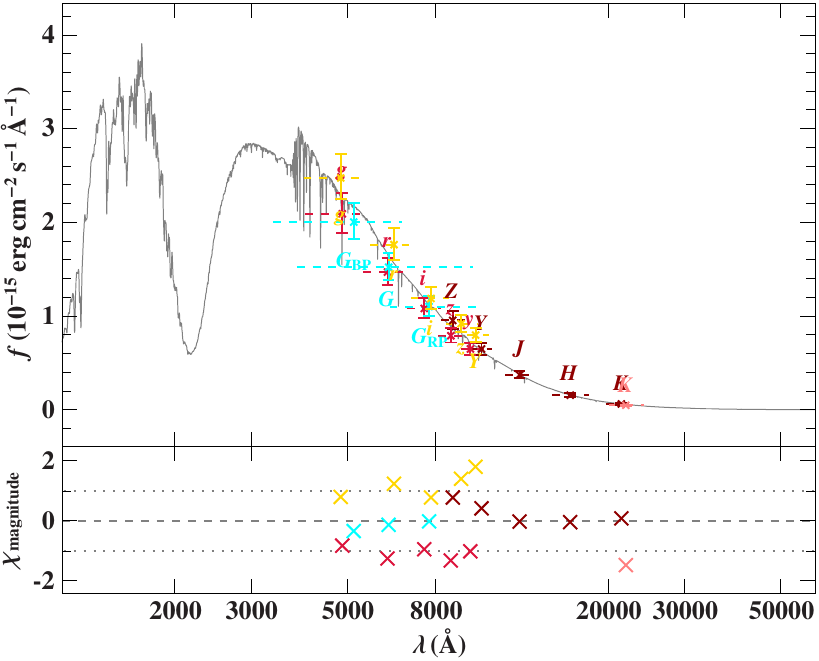}
    \caption{Spectral energy distribution fitting for OGLE-BLAP-009. 
    The best-fit model spectrum is shown as a grey line. 
    Colored data points represent flux measurements from different photometric surveys:
    \textit{Gaia} \citep[cyan,][]{gaia_phot}, 
    DECaPS \citep[yellow,][]{DECaPS_DR1}, 
    Pan-STARRS \citep[red,][]{PS1_DR2},
    VISTA/VVV \citep[dark red,][]{VVV_DR4},
    and UKIDSS \citep[light red,][]{UKIDSS_gps6}. 
    }
    \label{SED_fit}
\end{figure}

\begin{table}
\caption{List of input parameters and results obtained from the SED.}
\label{SED_res}
\centering
\renewcommand{\arraystretch}{1.16}
\begin{tabular}{lr}
\hline\hline
Parameter & Value \\
\hline
Color excess $E(44-55)$ & $0.670 \pm 0.025$\,mag \\
Extinction parameter $R(55)$ (fixed) & $3.02$ \\
Angular diameter $\log(\Theta\,\mathrm{(rad)})$ & $-10.942 \pm 0.020$ \\
Parallax $\varpi$  & $0.46 \pm 0.04$\,mas \\ 
Effective temperature $T_{\mathrm{eff}}$ (prescribed) & $28,000 \pm 1500$\,K \\
Surface gravity $\log (g\,\mathrm{(cm\,s^{-2})})$ (prescribed) & $4.40 \pm 0.20$ \\
Helium abundance $\log(n(\textnormal{He}))$ (prescribed) & $-0.60 \pm 0.20$ \\
Radius $R = \Theta/(2\varpi)$ & $0.55 \pm 0.06$\,$R_\odot$ \\
Mass $M = g R^2/G$  & $0.28^{+0.19}_{-0.12}$\,$M_\odot$ \\
Luminosity $L/L_\odot = (R/R_\odot)^2(T_\mathrm{eff}/T_{\mathrm{eff},\odot})^4$ & $170^{+60}_{-50}$ \\
\hline
\end{tabular}
\end{table}

As a first photogeometric method in estimating the mass and radius of OGLE-BLAP-009, a spectral energy distribution (SED) fit was performed. This method is used to determine the angular diameter of the star and combine it with the \textit{Gaia} EDR3 parallax to derive the stellar radius. The luminosity and mass are then calculated, making use of the atmospheric parameters determined from spectroscopy. This method was described in detail by \cite{Heber2018}. 
The $T_{\textrm{eff}}$ and $\log(g)$ were fixed at $28,000\pm1500$~K and $4.40\pm0.20$ based on the results from spectroscopy which do not include measurements made during phases of compression and with uncertainties that accommodate the range of values caused by the pulsation cycle. 
Interstellar reddening was accounted for using the functions of \cite{Fitzpatrick2019}, where the colour excess $E$\,(44$-$55) was a free parameter and the extinction parameter $R(55)$ was fixed to the standard value of 3.02. 
Fig.~\ref{SED_fit} compares the best-fit model spectrum to the available photometry.  
The radius was then calculated using $R=\Theta/2\varpi$, where $\Theta$ is the angular diameter estimated from the photometry and $\varpi$ is the parallax measurement acquired from \textit{Gaia} EDR3 \citep{Gaia_EDR3}. 
The parallax uncertainty was inflated according to Eq.\ 16 of \cite{El-Badry2021} and a zero-point offset of $-0.044$\,mas was applied following \cite{Lindegren21}.
\citet{gaiadr3var} classified the star as a variable of short period from \textit{Gaia} DR3 photometry with Standard deviations of 0.091\,mag and 0.074\,mag in the BP and RP bands, respectively. 
The mass was then calculated using $M=g R^2/G$, where $g$ is the surface gravity and $G$ is the gravitational constant. These calculations resulted in a mass of $0.28_{-0.12}^{+0.19}$~M$_{\odot}$ and a radius of $0.55\pm0.06$~R$_{\odot}$. 
There is a large uncertainty on the mass in this case as a consequence of the uncertainty embedded in $\log(g)$ due to the radius change occurring during radial-mode pulsations.

Table~\ref{SED_res} shows a list of all input parameters and results from the SED. Strong reddening can be seen in the photometry, due the location of OGLE-BLAP-009 in the Galactic disk. The estimated luminosity using the $T_{\textrm{eff}}$ input and derived radius  corresponds to $\log(L/L_{\odot}) = 2.25_{-0.20}^{+0.27}$. This value is consistent with the derived luminosity from \cite{BLAPDISC2017} for an approximately 0.30~M$_{\odot}$ hydrogen shell-burning star with a degenerate helium-core. 

\subsection{Pulsation characteristics} \label{pulse_char} 
\begin{figure}
	\includegraphics[width=\columnwidth]{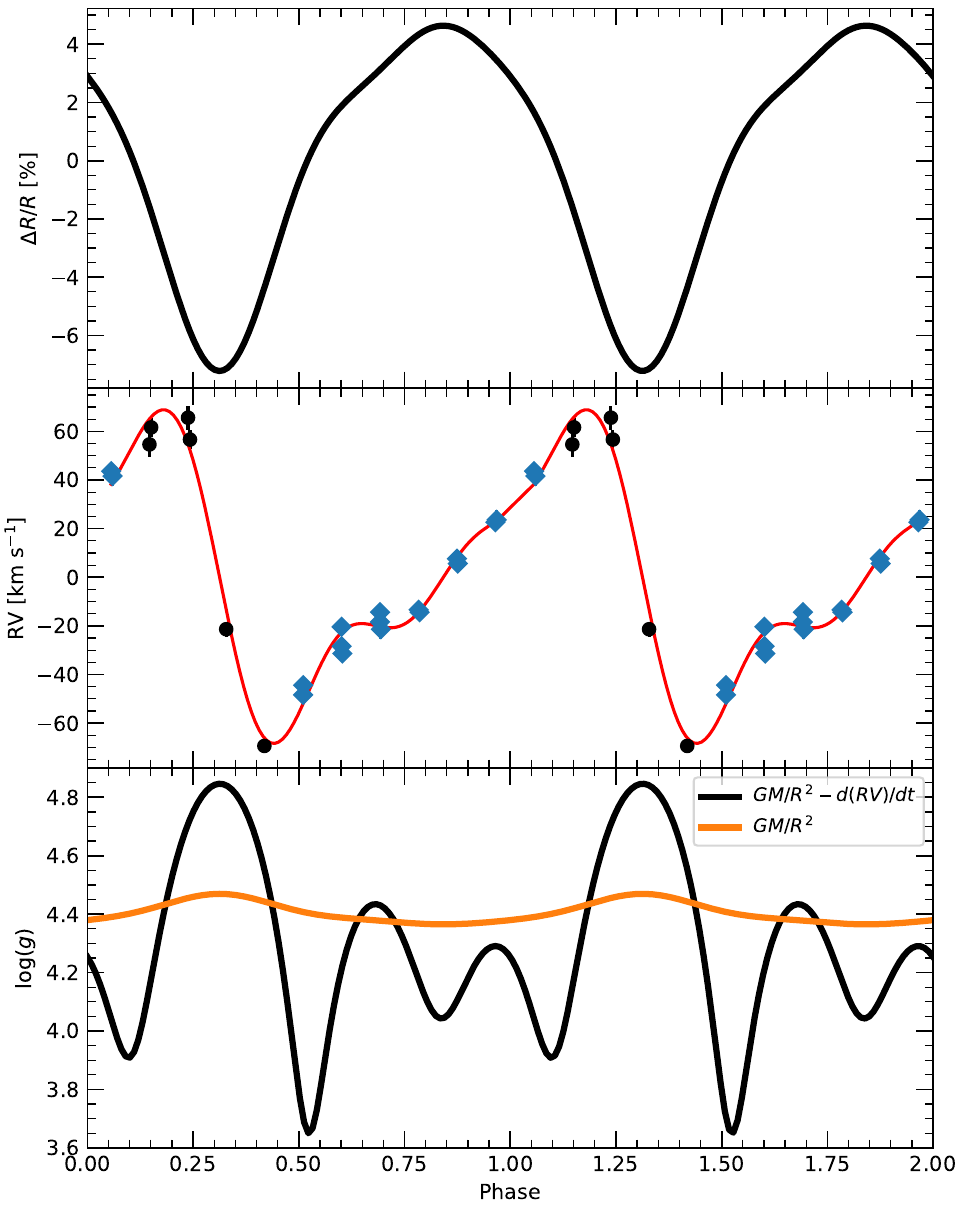}
    \caption{Pulsation properties calculated using a variant of the Baade-Wesselink method. Top panel: fractional radius change calculated using the integral of the $RV$ fitting (Eq.~\ref{eq2}), expressed as a percent of the radius derived from the SED. Second panel: $RV$ data points from Fig.~\ref{fig:spectroscopy} with the red line corresponding to the fitting model from Eq.~\ref{eq1}. Bottom panel: acceleration calculated using the derivative of the $RV$ fitting (black line) and the change in surface gravity calculated from the radius change (orange line).}
    \label{fig4}
\end{figure}

In order to explore the pulsation characteristics of OGLE-BLAP-009, we utilized a variant of the Baade-Wesselink method, which is frequently applied to radial-mode pulsators to calculate stellar parameters \citep{Baade_1926, Wesselink_1946}. We make use of the phase-folded $RV$ curve fitting from Eq.~\ref{eq1}. Using this model, we integrate and differentiate over the pulsation cycle to, respectively, find the radius change and surface gravity variations due to atmospheric acceleration. Fig.~\ref{fig4} displays the results of these calculations. Here, the top panel is the integral of the $RV$ curve (Eq.~\ref{eq2}) represented as a fractional radius change ($\Delta R/R$) relative to the radius derived from the SED and expressed as a percent, such that a $0\%$ variation occurs at 0.55~R$_{\odot}$. The bottom panel shows the $\log$ of the superposition of the acceleration at the stellar surface and the underlying {\logg} term ($\log(GM/R^2 - d(RV)/dt)$) found from the derivative of the $RV$ curve (black line), along with the change in surface gravity calculated from the change in radius alone ($\log(GM/R^{2})$) using the radius and mass found from the SED calculation (orange line). 

When integrating Eq.~\ref{eq1} over the pulsation cycle, it is necessary to adopt a projection factor to convert the measured radial velocities into true pulsation velocities. This factor takes into account a geometric projection due to limb darkening, a correction for the velocity gradient between the line forming regions and the photosphere, and relative motions between the gas and optical layers in the photosphere (see \citealt{projection_2007,projection_2013}). Following the works of \cite{projection_effect} and \cite{Barlow_2010} we used a projection factor of $p = 1.4$ which was shown to be appropriate for sdB-type stars and where limb darkening is the main contribution to this value. We find that using Eq.~\ref{eq2} the amplitude of the radius change over the pulsation cycle of OGLE-BLAP-009 is approximately 46,000 km.
\begin{equation} \label{eq2}
    \Delta R = -p \int_{t_{0}}^{t_{f}} \dot{r} dt 
\end{equation}
In terms of the radius found using the SED fitting, this is a fractional radius change ($\Delta R/R$) of about 12\%. The phase at minimum radius corresponds to maximum $T_{\textrm{eff}}$, $\log(g)$ and the photometric dip mentioned in Sec.~\ref{sec:Photometry}, all of which occur at phase 0.31. The initial flux peak occurs just before minimum radius at phase 0.27.

Differentiating Eq.~\ref{eq1} shows the acceleration at the surface due to the velocity rate of change about the underlying surface gravity ($\log(GM/R^2 - dRV/dt)$). The peak of this curve is in agreement with the peak surface gravity as seen from spectroscopy (Fig.~\ref{fig:spectroscopy}). Note that the second peak in the observed $\log(g)$, as seen at phase 0.70 in the bottom panel of Fig.~\ref{fig:spectroscopy} is also present in the derived surface acceleration. This may be due to the change in velocity of the photosphere as it approaches its maximum radius, which can be seen as a flattening of the $RV$ curve at this phase. To calculate the change of surface gravity coming from the radius fluctuation we use the mass derived from the SED and plot $\log(g)$ as a function of $\Delta R$. This resulted in an additional $\log(g)$ change of 0.12 which is a less significant contribution to the overall change in spectroscopic $\log(g)$. Based on these results and the $\log(g)$ amplitude seen in the spectroscopic data, we conclude the predominant factor in this variation is the acceleration of the photosphere. This is an indication that hydrostatic modeling may lead to an improper treatment of the pressure broadening in the spectral lines and inaccurate $\log(g)$ measurements at phases of minimum radius.

\section{Discussion}
\subsection{Stellar evolution}\label{sec:Stellar_Evolution}

\begin{figure}
	\includegraphics[width=\columnwidth]{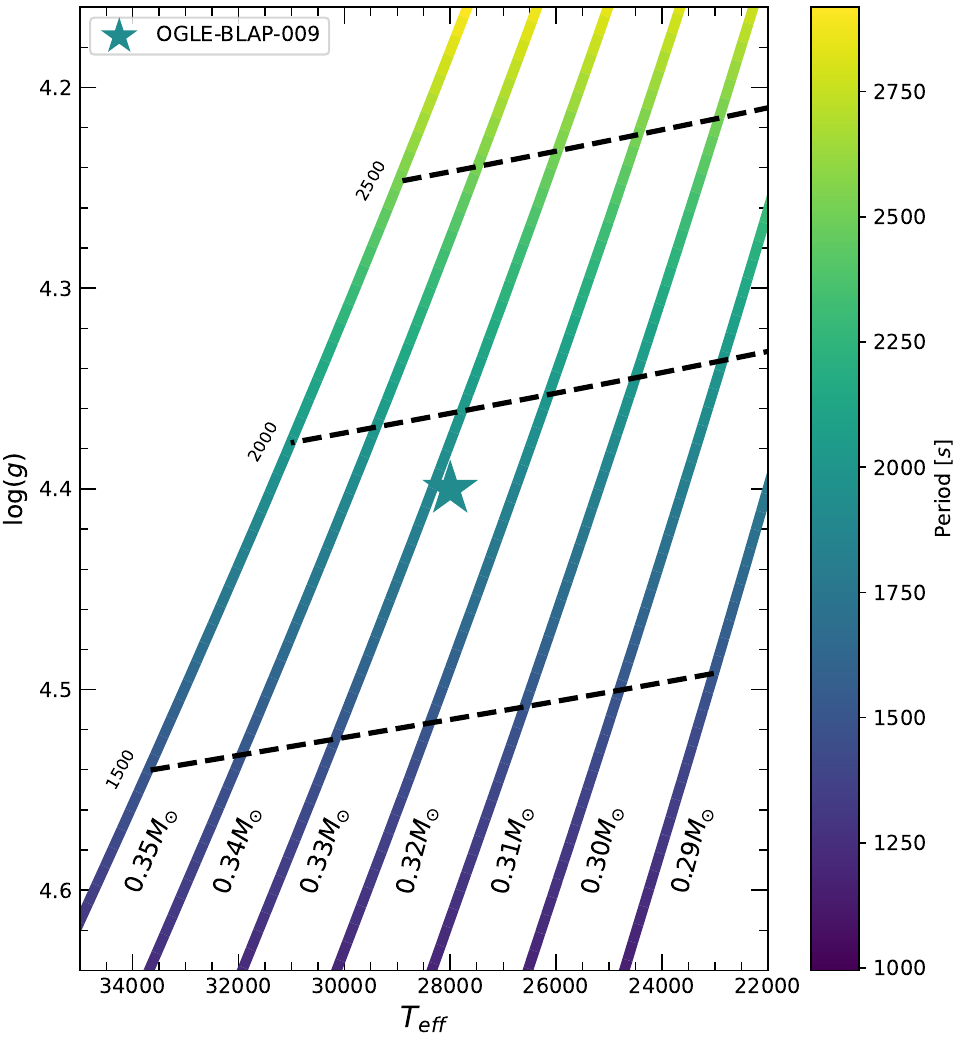} 
    \caption{MESA evolution tracks in $T_{\textrm{eff}}$ - $\log(g)$ space for He-core pre-WDs of masses between 0.29~$\textrm{M}_{\odot}$ and 0.35~$\textrm{M}_{\odot}$ in steps of 0.01~M$_{\odot}$. The tracks are color-mapped to the first-overtone radial-mode pulsation period calculated using GYRE along each step in the models' evolution. Period contours are also included as dashed black lines, which connect points of equal period across the models. OGLE-BLAP-009 is plotted using the mean $T_{\textrm{eff}}$ - $\log(g)$ acquired from spectroscopy and color-mapped to the scale of pulsation periods with an observed period of $\approx$1916.12~s.}
    \label{mesa}
\end{figure}

In order to apply a secondary, independent method of estimating the mass and radius of OGLE-BLAP-009 and to place a constraint on its evolutionary history, a grid of theoretical models for He-core pre-WDs of masses between 0.29 - 0.35~$\textrm{M}_{\odot}$ were computed using MESA \citep{mesa2011,mesa2013,mesa2015,mesa2018, mesa2019}. Additionally, linear adiabatic pulsation periods for both the fundamental and first-overtone radial-modes were calculated for each step along these tracks using GYRE \citep{gyre}. Fig.~\ref{mesa} shows OGLE-BLAP-009 plotted along with these models using the $T_{\textrm{eff}}$ and $\log(g)$ from Sec.~\ref{sec:SED} with the period color-mapped to the scale of first-overtone periods calculated using GYRE. The $T_{\textrm{eff}}$ and $\log(g)$ of OGLE-BLAP-009 best matches that of a 0.33 M$_{\odot}$ He-core pre-WD. The GYRE period for the fundamental mode pulsation at this model's closest matching point is 2402.60~s and 1900.60~s for the first-overtone. Therefore, the observed period of $\approx$1916.12~s is best described by a first-overtone radial-mode pulsation using these models (see Fig.~\ref{fig:A1}).

To calculate the radius and mass, a dimensionless frequency factor $f=\omega / \omega_{\rm dyn}$ was calculated for each point along the models' tracks, where $\omega$ is the mode frequency and $\omega_{\rm dyn} \equiv \sqrt{GM/R^3}$ is the stellar dynamical frequency. These factors from the best matching model are $f\approx3.42$ for the fundamental radial-mode and $f\approx4.31$ for the first-overtone radial-mode. If the observed pulsation frequency corresponds to a radial-mode of frequency $\omega$, then the radius and mass of the star can be calculated using the frequency factor $f$ for that mode according to the following.
\begin{equation}\label{eq3}
    R=\frac{10^{\langle{\log(g)\rangle}}f^{2}}{\omega^{2}}
\end{equation}
\begin{equation}\label{eq4}
    M=\frac{10^{\langle{\log(g)\rangle}}R^{2}}{G}=\frac{10^{\langle{3\log(g)\rangle}}f^{4}}{G\omega^{4}}
\end{equation}
Using Eqs.~\ref{eq3} \& \ref{eq4} and the average surface gravity from Sec.~\ref{sec:Spectroscopy}, we calculate the mass and radius for an assumed fundamental radial-mode pulsation to be $0.14\pm0.06$~M$_{\odot}$ and $0.39\pm0.07$~R$_{\odot}$. For an assumed first-overtone pulsation we calculate $0.36\pm 0.14$~M$_{\odot}$ to be the mass and $0.63\pm 0.11$~R$_{\odot}$ to be the radius. We find that the results calculated from the dynamical frequency of the first-overtone show agreement with a low-mass He-core pre-WD which resembles the radius and mass found using the SED and \textit{Gaia} parallax when considering the uncertainty caused by the changing photosphere. The period also closely resembles that of a predicted first-overtone pulsation at the mean $T_{\textrm{eff}}$ and $\log(g)$ that we calculate from spectroscopy. The mass and radius calculated from this model is also more consistent with the predicted He-core pre-WD mass of 0.33~M$_{\odot}$. Although, due to uncertainty in $\log(g)$ the fundamental mode is not fully excluded. However, these calculations for OGLE-BLAP-009 clearly show consistency with a low-mass He-core pre-WD of $\approx0.30$~M$_{\odot}$. 

\subsection{Stellar pulsations} \label{sec:Stellar_Pulsations}
The pulsation features of OGLE-BLAP-009 are similar to the pulsating extreme helium stars V652 Her and BX Cir \citep{Jeffery_2015,Woolf_2002}. \cite{Jeffery_2015} conducted a line-by-line spectral analysis of V652 Her and found that the large radial-mode pulsations resulted in extreme compression of the photosphere at minimum radius. At this phase, the acceleration of individual absorption lines that are formed deeper in the stellar interior was shown to initiate before those formed closer to the surface. Their findings suggest that a shock wave propagates through the photosphere as the pulsation mechanism is triggered \citep{Jeffery2022, Jeffery_etal_2022}. Due to the short pulsation period of OGLE-BLAP-009 and the lack of spectra covering multiple phases of radius minimum, we do not attempt to resolve such variations in the phase of initial acceleration of the individual absorption lines. However, based on the rapid expansion and contraction of this star's radius, similar effects may be likely to occur. 

The dip in flux occurring at maximum light may be a photometric indication of such a shock since it is present at the phase of minimum radius and stalls the flux decrease that occurs from radius expansion. The hydrodynamic pulsation models developed by \cite{Jeffery_etal_2022}, show that the flux from the immediate subsurface layers peaks shortly before the flux from the photosphere. This may be a reason for this photometric dip in flux and the phase delay between the first flux peak and minimum radius. The BLAP discovered by \citet{HD13_2022}, which has a 32.37 minute pulsation period also showed a similar feature in TESS 20-$s$ cadence data occurring just after maximum flux. It was speculated by \citet{monoperiod} that this double-peak feature could be the result of one peak corresponding to maximum temperature and the other maximum radius. This does not appear to be the case for OGLE-BLAP-009, as our phase-resolved results show that the point of maximum temperature occurs at the phase of minimum radius (see Sec.~\ref{pulse_char}). \citet{Macfarlane_2017} identified that this feature is similar to those found in ‘Bump’ Cepheid variables and is likely due to an exposure of the ionization shock front that is driving the pulsation as a result of the resonance between the fundamental and second-overtone pulsation modes. It is notable that these BLAPs which share this feature have a very similar pulsation period. This may be an indication that each of them share similar fundamental properties which result in the photometric presence of this feature.

The rapid variability in the observable properties of BLAPs creates the necessity for phase-resolved spectroscopy and proper spectral modeling. Both \citet{Jeffery_2015} and \citet{Woolf_2002} emphasize the need for hydrodynamic models to best recover the true atmospheric properties of high amplitude radially pulsating stars that are undergoing rapid changes of their radius. \citet{Jeffery2022} has investigated the impact of these models on V625 Her and we identify the need for similar models in analyzing BLAPs as well, based on the outlying $T_{\textrm{eff}}$ and $\log(g)$ observed at minimum radius and the strong surface acceleration. Additionally, spectroscopic observations covering multiple pulsation cycles may help to determine the true maximum values as the phase smearing and lack of pulsation peaks covered by the spectra leads to uncertainties at minimum radius. 

\subsection{Radial velocity monitoring}\label{sec:Radial_Velocity_Monitoring}
If the evolutionary status for OGLE-BLAP-009 is best described as a He-core pre-WD, then a previous interaction with a binary companion through common envelope evolution (CEE) or stable Roche lobe overflow (RLOF) was responsible for the mass loss of the BLAP progenitor \citep{Byrne_2021}. This binary evolution channel implies that there may be an observable companion that can be detected through long term radial velocity monitoring. We use the two Keck/ESI data sets to check if there is a variation in average $\langle RV \rangle$ about which the variability caused by stellar pulsations occurs. To quantify the mean radial velocity of each data set, both were phase folded and fitted individually with Eq.~\ref{eq1}, keeping each constant fixed excluding $a_{0}$, representing the average $\langle RV \rangle$, which was left as an open parameter. This resulted in $a_{0}=54.44\pm1.84$~km~s$^{-1}$ for the July 2020 data set and $a_{0}=49.72\pm2.88$~km~s$^{-1}$ for June 2021. An increase in uncertainty arises in the second data set since it covers less than one pulsation cycle. These results are the same within the uncertainty, however there is a range of $\approx$10 km~s$^{-1}$ in which a $RV$ shift may be obscured. 

\begin{figure}
	\includegraphics[width=\columnwidth]{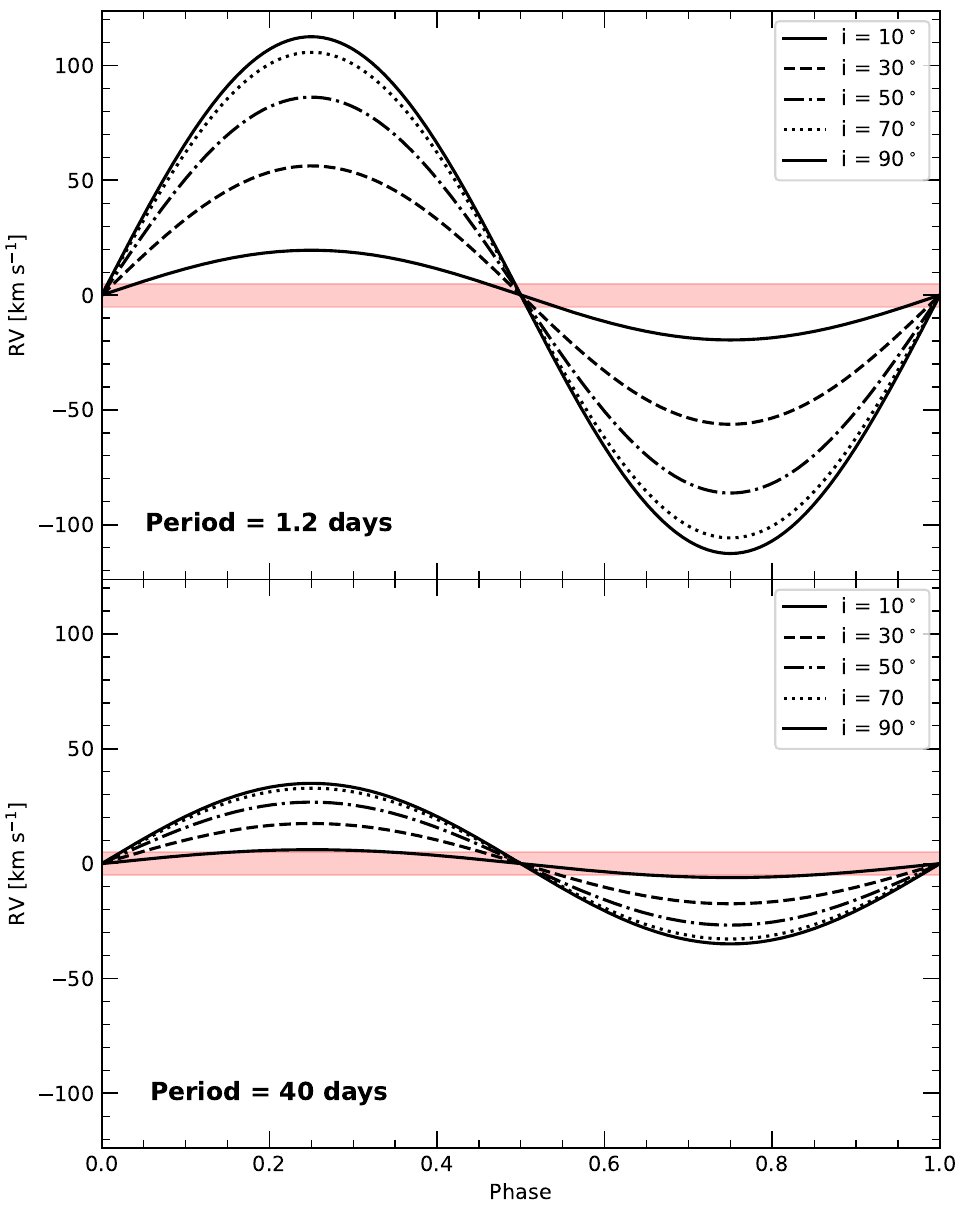}
    \caption{Radial velocity curve of possible binary orbits with various inclination angles, where OGLE-BLAP-009 is the secondary with an assumed mass of 0.34~M$_{\odot}$ and a low-mass main sequence primary of 0.50~M$_{\odot}$. The top panel is an assumed short period of 1.2 days and the bottom panel is a longer period assumption of 40 days. The average $\langle RV \rangle$ is assumed to be zero and the region of uncertainty from the average $\langle RV \rangle$ measurement of each Keck/ESI data set is shaded in red, where a shift may be obscured.}
    \label{binary}
\end{figure} 

\citet{Byrne_2021} conducted a binary population synthesis study to determine which formation channels produce low-mass pre-WDs that posses observable properties of BLAPs. They find that 75\% of their BLAP models emerge as a result of RLOF where a low-mass Main Sequence companion of $\approx$0.50M$_{\odot}$ is responsible for the mass stripping of the BLAP progenitor in most cases. There is a bimodal distribution in the final orbital periods of these models with a small peak at 1.2 days and a larger peak at 40 days. Assuming a He-core pre-WD mass of 0.34~M$_{\odot}$ from the best matching MESA model and a companion of 0.50~M$_{\odot}$, we simulate radial velocity curves for both of these orbital periods. Fig~\ref{binary} shows these curves plotted for inclination angles between 10$^{\circ}$ and 90$^{\circ}$ in steps of 20$^{\circ}$, centered about an average $\langle RV \rangle$ of zero with a region of uncertainty of $\pm$5~km~s$^{-1}$ shaded in red. Based off these assumptions, we find that a radial velocity shift for these binary orbits with inclination angles of <3$^{\circ}$ for the 1.2 day period and <10$^{\circ}$ for the 40 day period can be completely obscured. The probability of observing it at a phase within this obscured region decreases exponentially with inclination angle, down to an edge-on (90$^{\circ}$) orbit. In this orientation a total time spent with a radial velocity amplitude of less than 10~km~s$^{-1}$ is 3\% of the orbit for a 1.2 day period and 9\% of the orbit for a 40 day period. It is therefore unlikely that an average $\langle RV \rangle$ shift would be missed, especially in the case of a short orbital period.

Using $O-C$ analysis of light-travel-time effects, \citet{HD13_2022} reported an approximate 23.08433 day period of a BLAP around a late B-type main-sequence star in the binary system HD 133729. This constitutes the first BLAP discovered which lies in a confirmed binary system. If this BLAP shares a common origin with OGLE-BLAP-009, then it is possible that a binary orbital period would fall closer to the longer-period $RV$ amplitude distributions, leading to a slightly increased probability of a non-detection of the companion. However, it is still unlikely that an average $\langle RV \rangle$ shift from an orbital period within this range would be missed at inclinations above 10\degree. Additionally, \citet{HD13_2022} show a decrease in the photometric pulsation amplitude of the BLAP due to light dilution from the bright companion, which is not present in OGLE-BLAP-009. Therefore, if a companion exists, it is likely a dim companion such as the one reported by \citet{tmts_blap} from TMTS-BLAP1 results, and a longer orbital period is likely. This suggests that a proper $O-C$ analysis should be performed for OGLE-BLAP-009 and others like it in order to search for a binary companion. 

\subsection{Kinematics}\label{sec:Kinematics}

\begin{figure}
	\includegraphics[width=\columnwidth]{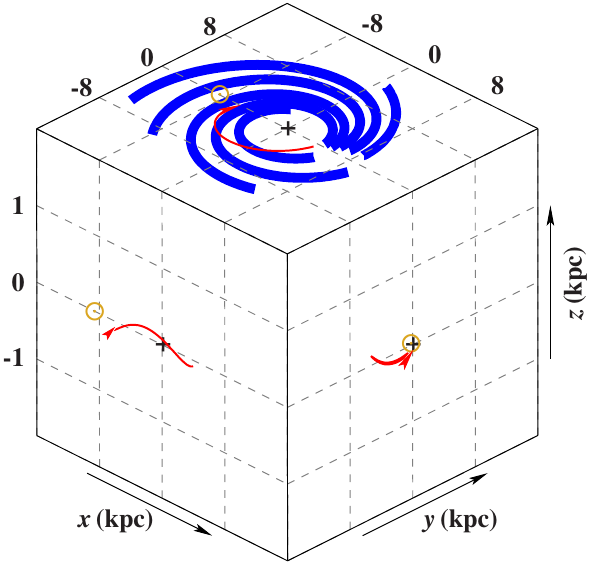}
    \caption{3D Galactic orbit of OGLE-BLAP-009. The red line indicates the orbit of OGLE-BLAP-009 relative to the galactic plane (blue), Galactic center (plus) and Sun (yellow).}
    \label{orbit}
\end{figure}

\begin{figure}
	\includegraphics[width=\columnwidth]{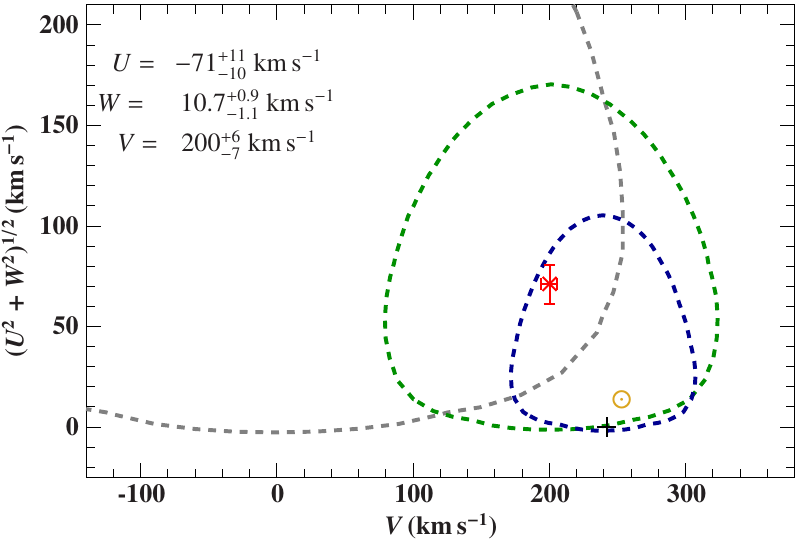}
    \caption{Toomre diagram for OGLE-BLAP-009 where the x-axis is the velocity in the direction of Galactic rotation ($V$), and the y-axis are the tangential velocities in the direction of the Galactic Center ($U$) and Galactic North Pole ($W$), summed in quadrature. The blue, green, and grey dashed lines respectively represent the 2$\sigma$ contours for the thin disk, thick disk and halo, based on data from \citet{Robin2003}.
    }
    \label{Toomre}
\end{figure}

The Galactic trajectory of OGLE-BLAP-009 was calculated as described in \citet{irrgang2013} using their Galactic model potential I. This showed a usual orbit for a thin disk object with origins in the inner part of the disk. Fig.~\ref{orbit} shows its 3D galactic orbit, relative to the galactic center and Sun. This star stays within $\approx$~100 pc of the Galactic plane in the z direction, which is consistent with a thin disk object. However, the orbit is eccentric as can be seen in the x-y plane of Fig.~\ref{orbit}. Fig.~\ref{Toomre} shows a Toomre diagram for the space velocity of OGLE-BLAP-009. The eccentric orbit leads to it being settled on the bordering region between a thin disk (blue) and thick disk object (green). Tracing the orbit back 70 Myr, shows it originating from the inner part of the disk, $\approx$4 kpc from the center. If the formation of OGLE-BLAP-009 occured in this region, this could be an explanation for the increased metallicity seen in Sec.~\ref{sec:Metal Abundance}. The high metallicity in combination with the strong nitrogen enrichment points to an intermediate mass main sequence progenitor. 

\section{Summary and Conclusion}

We present follow up photometry and time-series spectroscopy of the blue large-amplitude pulsator OGLE-BLAP-009 in order to estimate the radius and mass of the star and place constraints on its evolutionary origins through comparison of its observed properties with theoretical models of He-core pre-WDs. We find large variability in flux, $RV$, $T_{\textrm{eff}}$ and $\log(g)$ as a result of the rapid change in radius of about 12\%. We see no clear signs of a binary companion through photometry or through spectroscopy via an average $\langle RV \rangle$ shift between Keck/ESI data sets taken approximately one year apart. However, due to the uncertainty in $\langle RV \rangle$ we cannot rule out any binary orbits that would result in a radial velocity shift of less than 10 km~s$^{-1}$. 

A spectral energy distribution fitting of available photometric flux measurements to hot subdwarf models, combined with the \textit{Gaia} parallax and using the $T_{\textrm{eff}}$ and $\log(g)$ derived from spectroscopy resulted in a mass of 0.28$_{-0.12}^{+0.19}\,\textrm{M}_{\odot}$ and radius of $0.55\pm0.06\,\textrm{R}_{\odot}$. Placing OGLE-BLAP-009 on a $T_{\textrm{eff}}$ -- $\log(g)$ diagram along with evolution tracks of He-core pre-WDs computed using MESA showed a close match with the model of mass 0.33~M$_{\odot}$. Predicted periods of fundamental and first-overtone radial-mode pulsations were calculated for each model using GYRE and the observed period best matched that of the first-overtone radial-mode. Calculating the mass and radius using a dynamical frequency factor of the closest matching model in period, $T_{\textrm{eff}}$ and $\log(g)$ for the first-overtone mode resulted in a mass of $0.36\pm 0.14\,M_{\odot}$ and radius of $0.63\pm 0.11\,R_{\odot}$ which is consistent with the SED results. 

Our study indicates that the observed properties of OGLE-BLAP-009 can best be explained by a low-mass He-core pre-WD of $\approx$0.30~M$_{\odot}$. There are uncertainties in the mass estimates due to the change in surface gravity caused by stellar pulsations, but the results show a clear constraint to an evolutionary scenario that results in a low mass object of less than 0.50~M$_{\odot}$. Further photometric analysis of the observed light curve peaks over time may help to confirm if OGLE-BLAP-009 is, in fact, in a binary orbit. Using $O-C$ analysis of pulse timing measurements can reveal orbital reflex motion due to light travel time effects. This has proven useful in the study of the only two BLAPs discovered to be in a binary system, HD 133729 and TMTS-BLAP1 \citep{HD13_2022,tmts_blap}. Confirming binary orbits in BLAPs and high-gravity BLAPs will help in constraining evolutionary channels that lead to their formation. The results presented here show that this method of analysis will prove useful if applied to other BLAPs in order to characteristics their observable properties and attempt to place constraints on their evolutionary history. Additional spectroscopic observations are currently underway in order to apply this analysis to a statistically significant sample and determine if the class of BLAPs as a whole share similar characteristics and evolutionary status. 

\section*{Acknowledgements}

We thank Andreas Irrgang for the development of the spectrum and SED-fitting tools and his contributions to the model atmosphere grids. BNB acknowledges support from National Science Foundation award AST 1812874.  TK acknowledges support from the National Science Foundation through grant AST \#2107982, from NASA through grant 80NSSC22K0338 and from STScI through grant HST-GO-16659.002-A. This work was supported, in part, by the National Science Foundation through grant PHY-1748958, and by meetings supported by the Gordon and Betty Moore Foundation through grant GBMF5076. Co-funded by the European Union (ERC, CompactBINARIES, 101078773). Views and opinions expressed are however those of the author(s) only and do not necessarily reflect those of the European Union or the European Research Council. Neither the European Union nor the granting authority can be held responsible for them.

This work has made use of data from the European Space Agency (ESA) mission
{\it Gaia} (\url{https://www.cosmos.esa.int/gaia}), processed by the {\it Gaia}
Data Processing and Analysis Consortium (DPAC,
\url{https://www.cosmos.esa.int/web/gaia/dpac/consortium}). Funding for the DPAC
has been provided by national institutions, in particular the institutions
participating in the {\it Gaia} Multilateral Agreement.

\section*{Data Availability}

The spectroscopic data taken with Keck/LRIS and Keck/ESI can be downloaded from the Keck archive. The photometric data used in this paper will be provided upon request.



\bibliographystyle{mnras}
\bibliography{BIB.bib} 



\appendix
\section{Additional Figures}

\begin{figure*}
    \centering
    \includegraphics[width=\textwidth]{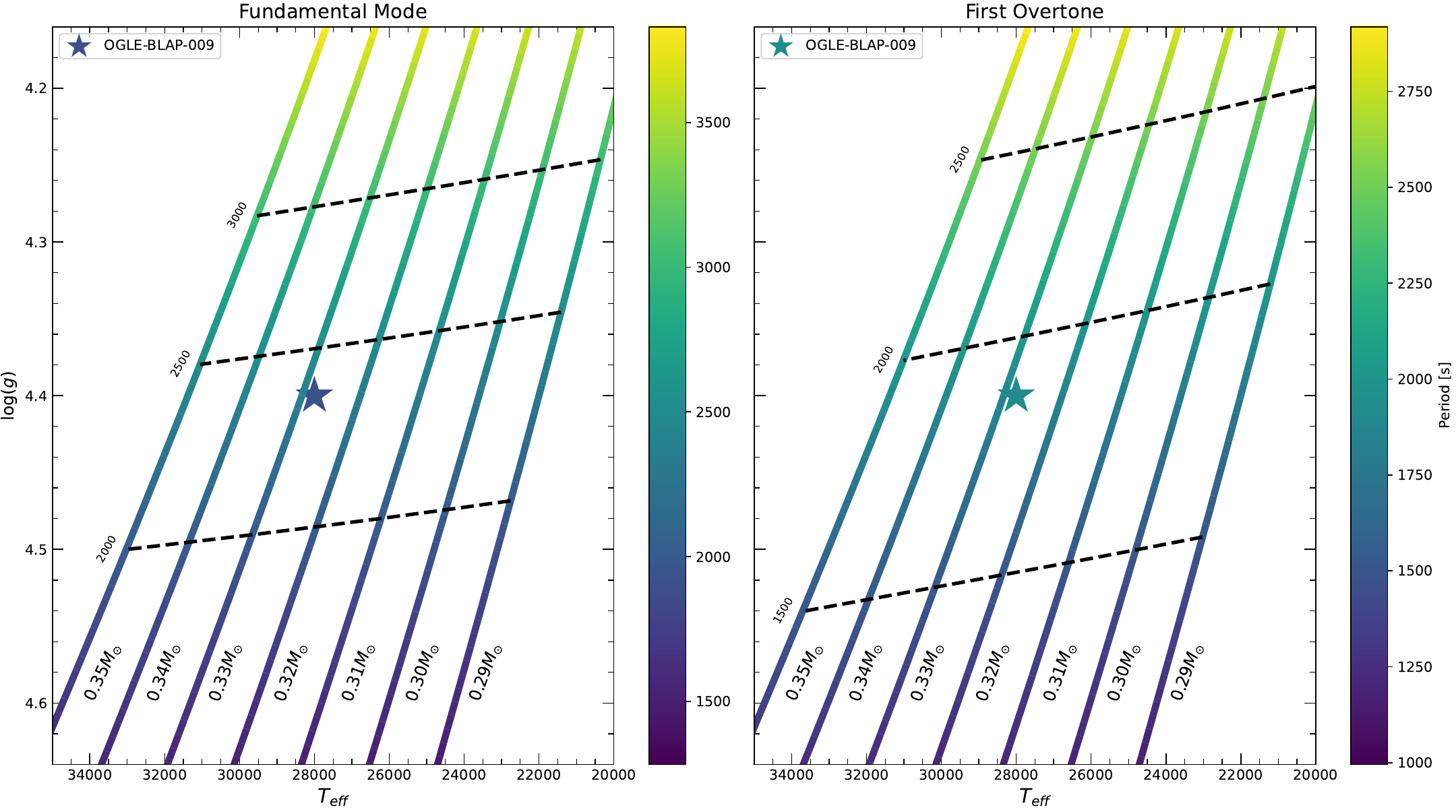}
    \caption{A comparison of the GYRE fundamental and first-overtone radial-mode pulsation periods matched to the observed period of OGLE-BLAP-009 at $\approx$1916.12~s. Left panel: MESA He-core pre-WD $T_{\textrm{eff}}$ -- $\log(g)$ models color-mapped to the fundamental radial-mode GYRE periods. Right panel: the same MESA models color-mapped to the first-overtone radial-mode GYRE periods. Period contours are also included for each panel as dashed black lines, which connect points of equal period across the models.}
    \label{fig:A1}
\end{figure*}

Fig.~\ref{fig:A1} shows a comparison between the pulsation period match of OGLE-BLAP-009 to the periods calculated for the fundamental and first-overtone radial-modes. The period of OGLE-BLAP-009 more closely matches the predicted period for a first-overtone pulsation at the mean $T_{\textrm{eff}}$ and $\log(g)$ derived from spectroscopy. Fig.~\ref{fig:full_spectrum} shows a full comparison between the co-added ESI spectrum and our best-fit \textsc{Tlusty}/\textsc{Synspec} model.

\captionsetup[ContinuedFloat]{labelformat=continued}
\begin{figure*}
\centering
\includegraphics[width=23.2cm,angle=90,page=1]{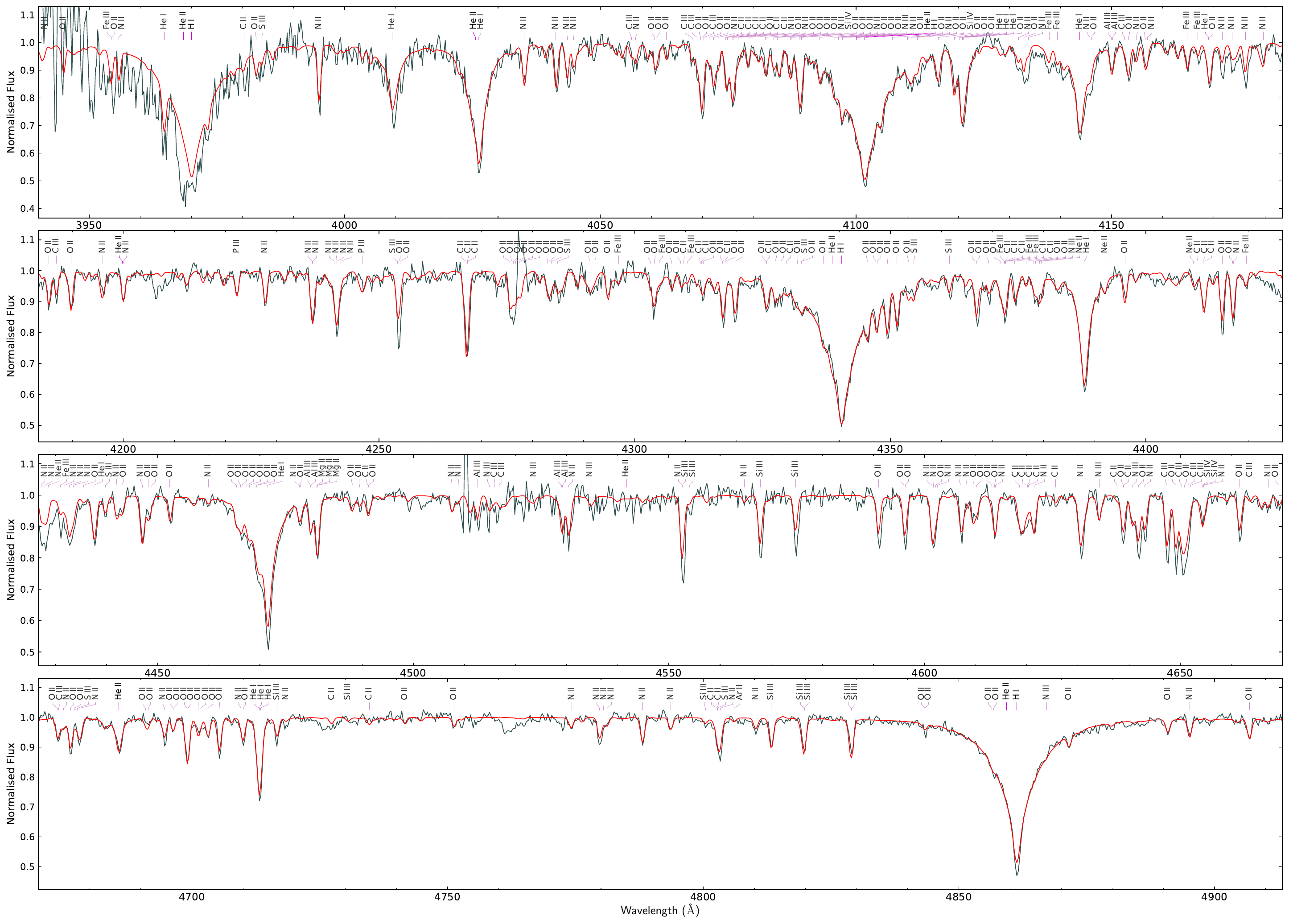}
\caption{Co-added ESI spectrum of OGLE-BLAP-009 (grey) and the final model (red).}
\label{fig:full_spectrum}
\end{figure*}
\begin{figure*}
\ContinuedFloat
\centering
\includegraphics[width=23.2cm,angle=90,page=2]{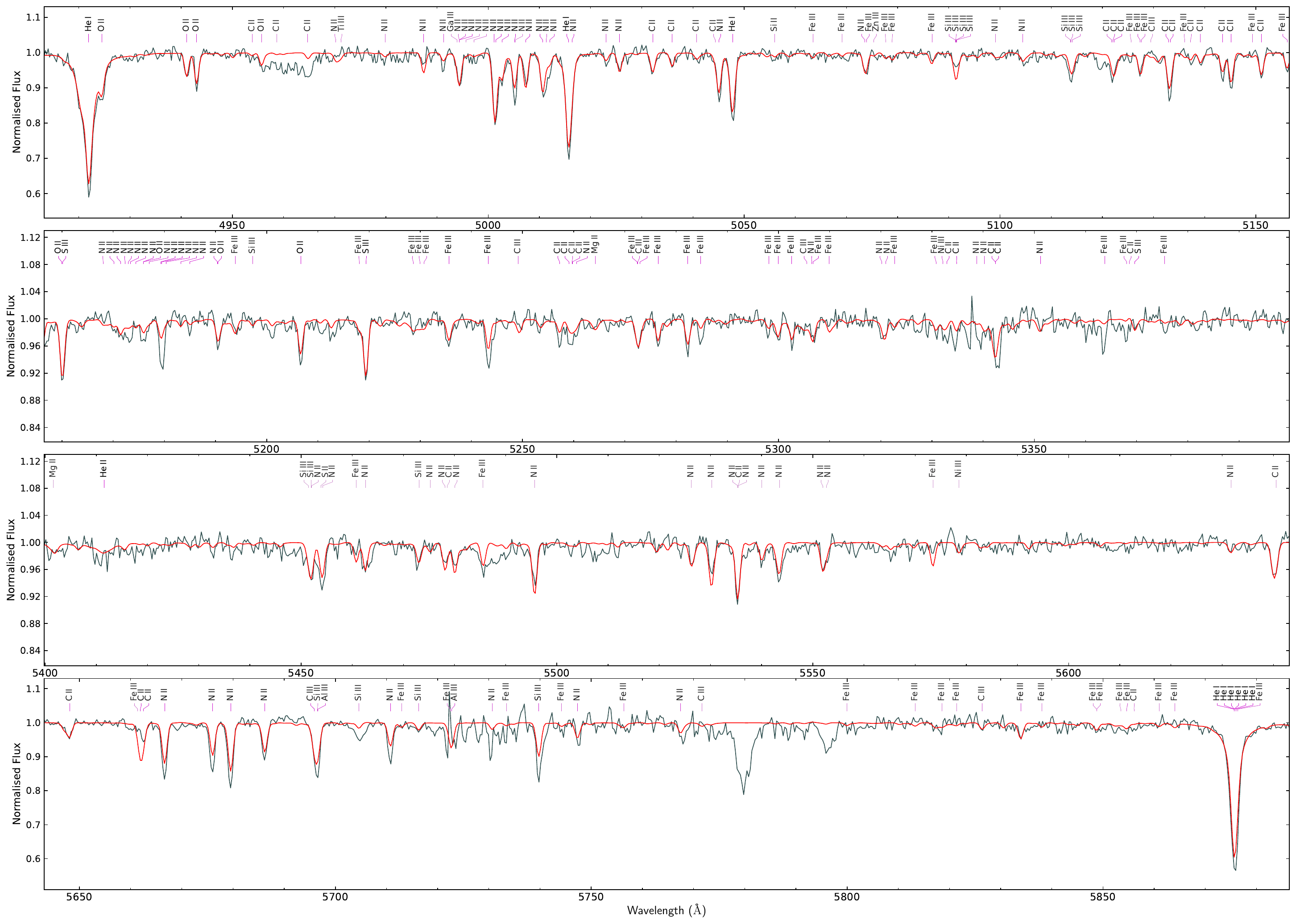}
\caption{Co-added ESI spectrum of OGLE-BLAP-009 (grey) and the final model (red).}
\end{figure*}
\begin{figure*}
\ContinuedFloat
\centering
\includegraphics[width=23.2cm,angle=90,page=3]{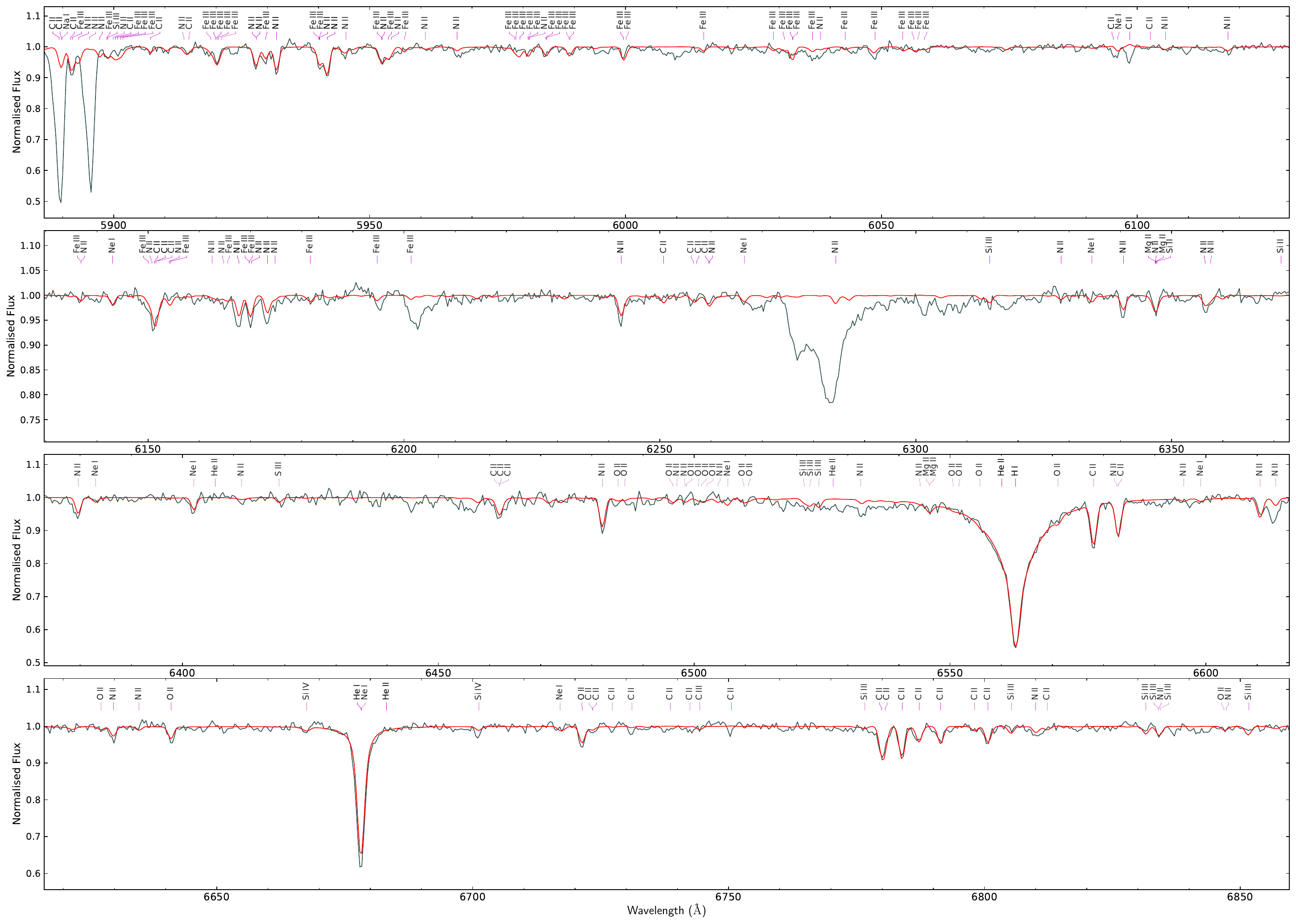}
\caption{Co-added ESI spectrum of OGLE-BLAP-009 (grey) and the final model (red).}
\end{figure*}

\section{Table of values from phase-resolved spectroscopy. }
\begin{table*}
	\centering
	\caption{List of values from phase-resolved spectroscopy. The instrument used and date of acquisition are available for each data point. Corresponding uncertainties for each measurement are included, which are statistically small, only on the order of 1-$\sigma$.}
	\label{tab:tabB1}
	\begin{tabular}{lcccccc} 
		\hline
		\hline
		Instrument & Date & Phase & $RV$ [km~s$^{-1}$] & $T_{\textrm{eff}}$ [K] & $\log(g)$ & $\log(y)$\\
		\hline
		ESI & July 2020 & 0.05633967 & $97\pm3$  & - & - & -\\
		ESI & June 2021 & 0.05957654 & $95\pm4$  & - & - & -\\
        LRIS & Sept. 2021 & 0.06987998 & - & $27290\pm633$ & $4.34\pm0.08$ & $-0.68\pm0.12$\\
		ESI & July 2020 & 0.14683326 & $108\pm5$  & - & - & -\\
		ESI & June 2021 & 0.15084299 & $115\pm4$  & - & - & -\\
        LRIS & Sept. 2021 & 0.15184787 & - & $27937\pm612$ & $4.43\pm0.08$ & $-0.73\pm0.15$\\
        LRIS & Sept. 2021 & 0.23381577 & - & $30209\pm1219$ & $4.51\pm0.09$ & $-0.47\pm0.11$\\
        ESI & July 2020 & 0.23809338 & $119\pm5$  & - & - & -\\
        ESI & June 2021 & 0.24269605 & $110\pm4$  & - & - & -\\
        LRIS & Sept. 2021 & 0.31555822 & - & $36660\pm558$ & $4.80\pm0.05$ & $-0.74\pm0.13$\\
        ESI & July 2020 & 0.32831641 & $32\pm3$  & - & - & -\\
        LRIS & Sept. 2021 & 0.39752612 & - & $32959\pm1250$ & $4.50\pm0.10$ & $-0.63\pm0.11$\\
        ESI & July 2020 & 0.4185399 & $-16\pm2$  & - & - & -\\
        LRIS & Sept. 2021 & 0.47949401 & - & $32226\pm865$ & $4.30\pm0.05$ & $-0.56\pm0.08$\\
        ESI & July 2020 & 0.51017629 & $5\pm2$  &- & - & -\\
        ESI & July 2020 & 0.5103862 & $9\pm2$  & - & - & -\\
        LRIS & Sept. 2021 & 0.56146188 & - & $30207\pm676$ & $4.27\pm0.08$ & $-0.66\pm0.16$\\
        ESI & July 2020 & 0.60115032 & $33\pm2$  &- & - & -\\
        ESI & July 2020 & 0.60121095 & $25\pm2$  &- & - & -\\
        ESI & June 2021 & 0.60306264 & $22\pm2$  & - & - & -\\
        LRIS & Sept. 2021 & 0.6437003 & - & $28357\pm868$ & $4.47\pm0.08$ & $-0.78\pm0.14$\\
        LRIS & Sept. 2021 & 0.65976997 & - & $27114\pm752$ & $4.26\pm0.09$ & $-0.53\pm0.12$\\
        ESI & July 2020 & 0.69197509 & $35\pm3$  & - & - & -\\
        ESI & July 2020 & 0.692185 & $39\pm3$ & - & - & -\\
        ESI & June 2021 & 0.69410406 & $32\pm4$  & - & - & -\\
        LRIS & Sept. 2021 & 0.72539766 & - & $27775\pm679$ & $4.51\pm0.08$ & $-0.67\pm0.10$\\
        LRIS & Sept. 2021 & 0.74173786 & - & $27494\pm712$ & $4.51\pm0.08$ & $-0.72\pm0.12$\\
        ESI & July 2020 & 0.78350575 & $40\pm3$ & - & - & -\\
        ESI & June 2021 & 0.78568615 & $39\pm3$ & - & - & -\\
        LRIS & Sept. 2021 & 0.80736555 & - & $25792\pm551$ & $4.20\pm0.09$ & $-0.62\pm0.10$\\
        LRIS & Sept. 2021 & 0.82370576 & - & $28222\pm664$ & $4.29\pm0.07$ & $-0.59\pm0.11$\\
        ESI & July 2020 & 0.87399933 & $61\pm2$ & - & - & -\\
        ESI & June 2021 & 0.87591594 & $59\pm3$ & - & - & -\\
        LRIS & Sept. 2021 & 0.88933344 & - & $26473\pm715$ & $4.16\pm0.08$ & $-0.52\pm0.10$\\
        LRIS & Sept. 2021 & 0.90567367 & - & $26056\pm891$ & $4.18\pm0.07$ & $-0.50\pm0.10$\\
        ESI & July 2020 & 0.96503446 & $76\pm3$ & - & - & -\\
        ESI & June 2021 & 0.96799402 & $77\pm4$ & - & - & -\\
        LRIS & Sept. 2021 & 0.98764157 & - & $26311\pm567$ & $4.18\pm0.09$ & $-0.60\pm0.14$\\
		\hline
		\hline
	\end{tabular}
\end{table*}


\bsp	
\label{lastpage}
\end{document}